\documentclass[pdflatex,sn-mathphys-num]{sn-jnl}

\usepackage{graphicx}%
\usepackage{multirow}%
\usepackage{amsmath,amssymb,amsfonts}%
\usepackage{amsthm}%
\usepackage{mathrsfs}%
\usepackage[title]{appendix}%
\usepackage{xcolor}%
\usepackage{textcomp}%
\usepackage{manyfoot}%
\usepackage{booktabs}%
\usepackage{algorithm}%
\usepackage{algorithmicx}%
\usepackage{algpseudocode}%
\usepackage{listings}%

\theoremstyle{thmstyleone}%
\theoremstyle{thmstyletwo}%

\theoremstyle{thmstylethree}%

\begin{document}

\title[Article Title]{Foundations of GenIR}


\author[1]{\fnm{Qingyao} \sur{Ai}}\email{aiqy@tsinghua.edu.cn}
\equalcont{These authors contributed equally to this work.}

\author[1]{\fnm{Jingtao} \sur{Zhan}}\email{zhanjt20@mails.tsinghua.edu.cn}
\equalcont{These authors contributed equally to this work.}

\author[1]{\fnm{Yiqun} \sur{Liu}}\email{yiqunliu@tsinghua.edu.cn}

\affil[1]{\orgdiv{Dept. of Computer Science and Technology}, \orgname{Tsinghua University}, \city{Beijing}, \country{China}}




\abstract{The chapter discusses the foundational impact of modern generative AI models on information access (IA) systems. In contrast to traditional AI, the large-scale training and superior data modeling of generative AI models enable them to produce high-quality, human-like responses, which brings brand new opportunities for the development of IA paradigms. In this chapter, we identify and introduce two of them in details, i.e., information generation and information synthesis. Information generation allows AI to create tailored content addressing user needs directly, enhancing user experience with immediate, relevant outputs. Information synthesis leverages the ability of generative AI to integrate and reorganize existing information, providing grounded responses and mitigating issues like model hallucination, which is particularly valuable in scenarios requiring precision and external knowledge. This chapter delves into the foundational aspects of generative models, including architecture, scaling, and training, and discusses their applications in multi-modal scenarios. Additionally, it examines the retrieval-augmented generation paradigm and other methods for corpus modeling and understanding, demonstrating how generative AI can enhance information access systems. It also summarizes potential challenges and fruitful directions for future studies.}

\maketitle


The primary distinction between modern generative models and traditional AI techniques lies in their capability to generate complicated and high-quality output based on human instructions. 
As shown by many studies~\cite{gpt4,llm_survey,touvron2023llama}, modern generative AI models possess remarkable abilities to generate responses that closely mimic human interaction. 
General speaking, such impressive performance comes from their large-scale training collections and their advanced data modeling algorithms.
Their superior data understanding ability can benefit almost every components of existing information access systems, from document encoding and index construction, to query processing and relevance analysis, etc.
However, when talking about new opportunities or paradigms that are uniquely brought by the generative AI to information access, they can be broadly categorized in two directions.
The first one is to create content that directly addresses user's information needs. 
By understanding and taking user queries as input instructions, generative AI models are able to generate specific answers or products tailored to the individual's request. 
This direct approach to information generation can significantly enhance user experience by providing immediate and relevant responses.
The second direction is to leverage the advanced instruction-following capabilities of generative AI models to synthesize and recombine existing information in innovative ways. 
Generative AI such as large language models (LLMs) can take existing data and transform it into new, coherent pieces of information that may not have been explicitly outlined before. 
This ability to reinterpret and organize information opens up new possibilities for retrieval system design and applications.
Therefore, in this chapter, we discuss how generative AI models could help information access from two perspectives, namely \textit{information generation} and \textit{information synthesis}.

\section{Information Generation}\label{ch2-gen}


Information need is diverse and typically long-tail. Traditional information retrieval systems, such as search engines and recommendation platforms, are designed to present information that already exists. However, these systems often fall short when it comes to fulfilling the less common information needs. This is particularly evident in scenarios requiring creative creation, where users seek not just information but inspiration and novel ideas. The limitations of traditional information systems in addressing these unique demands have paved the way for the emergence of generative models, which hold the promise of creating new information that aligns closely with the long-tail information needs.

In recent years, generative models have made significant developments. For instance, ChatGPT can respond to user questions, Bing enhances its responses with retrieval-augmented generation, and Midjourney generate images based on user prompts, and recommendation systems generate personal contents for different users. The development is mainly driven by the capable model architectures, computational resources, and the large-scale internet data. These elements have facilitated the performance of generative models to new heights. With the continuous efforts on scaling up these elements, the model performance is still rapidly improving. Nowadays, generative models have gradually been integrated into various workflows and everyday life activities. 

In this section, we present the foundation of generative models.
This section is organized as follows. Section~\ref{sec:foundation_architecture} shows the efforts on designing the model architectures for large language models. Section~\ref{sec:foundation_scaling} discusses how scaling facilitates the development of generative models and its potential future. Section~\ref{sec:foundation_training} presents the different training stages of large language models. Finally, Section~\ref{sec:foundation_mm} introduces how large language models are used in multi-modal scenarios.

\subsection{Model Architecture}
\label{sec:foundation_architecture}

\begin{figure}[ht] 
	\centering
	\includegraphics[width=0.6\textwidth]{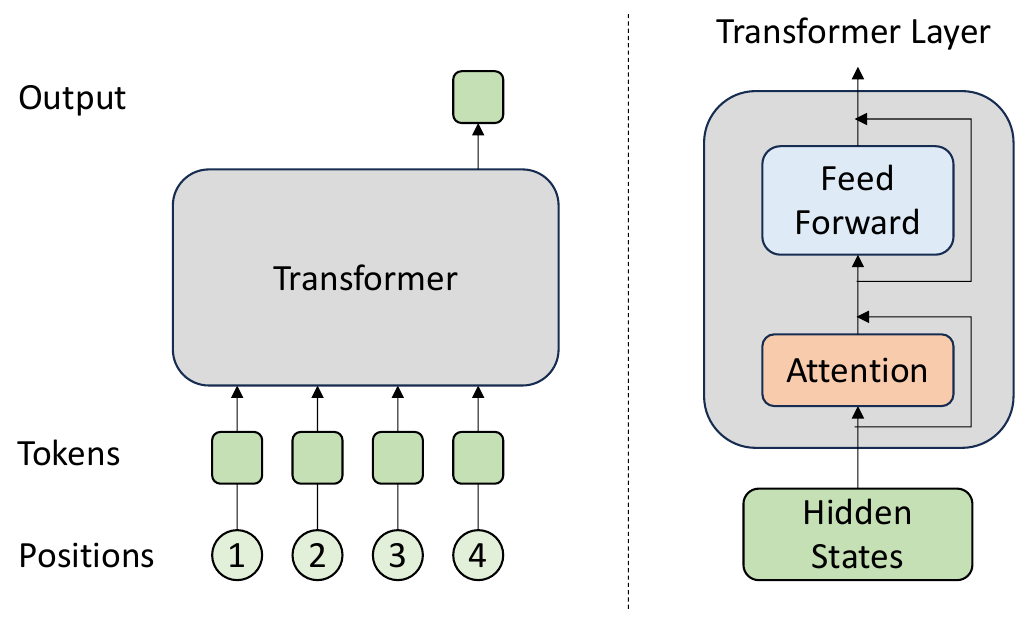} 
	\caption{Transformer architecture: the overview on the left and the illustration of one layer on the right~\cite{NIPS2017_attention}.}
	\label{fig:transformer}
\end{figure}

In different generation scenarios like ChatGPT or SoRA, Transformer~\cite{NIPS2017_attention} has emerged as the predominant model structure. It starts with an embedding layer, followed by multiple neural layers. Within each layer, an attention mechanism models the interactions between words, creating contextualized embeddings. The final decision on word generation probabilities is derived by comparing the output embedding with the vocabulary embeddings. We illustrate the model architecture in Figure~\ref{fig:transformer}.
Unlike traditional Recurrent Neural Networks~\cite{schuster1997bidirectional}, Transformers are capable of modeling long-distance interactions between words directly, which provides a more powerful representational capability. Numerous enhancements to the Transformer architecture have been proposed. In the following, we will explore various modifications to each component of the Transformer, highlighting the advancements that have further improved its efficacy and efficiency.

\subsubsection{Word Embedding}
Word embedding module is at the bottom of the Transformer architecture. Initially, a tokenizer breaks down a sentence into tokens, which the Word embedding module then maps into embeddings. These are combined with position embeddings and fed into subsequent neural layers. Recent research on large-scale language models has identified word embeddings as one of the main sources to training instability~\cite{zeng2022glm}. Particularly in the early stages of training, the gradients of word embeddings are often orders of magnitude larger than those of other parameters. To address this issue, \citet{le2023bloom} introduced a layer normalization immediately after the word embedding layer, stabilizing the distribution effectively. Besides, \citet{zeng2022glm} opted to scale down the gradients of the word embeddings by an order of magnitude to prevent substantial updates. Both approaches have been proven effective in stabilizing the training of language models at the 100 billion parameter scale. Yet, whether they are still effective for larger models remains to be investigated.

\subsubsection{Position Embedding}

Position embedding is essential for Transformer. Unlike RNNs that inherently process sequences in order, vanilla attention mechanism disregards the positional distances between words and Transformer has to rely on position embeddings for position modeling. Initially, Transformer~\cite{NIPS2017_attention} utilized Sinusoidal embeddings, a non-trainable form of position embedding that is added directly to word embeddings. Later, \citet{naccl19_BERT} introduced trainable position embeddings, which is initialized randomly and are updated through gradient descent during training. 
Subsequently, \citet{jmlr_T5} and \citet{press2021train} proposed relative positioning, where the attention mechanism incorporates biases based on the relative positions of words to better model varying distances. Recently, \citet{su2023enhanced} introduced the concept of rope position embedding, based on the principle that the dot product of vectors correlates with their magnitudes and the angles between them. By rotating vectors in space proportionally to their positions, this method naturally integrates positional information into attention scores. \citet{black2022gpt} has found that this approach outperforms trainable position embeddings.
Yet, these approaches may not work well when extrapolated to long sequences and more effective methods need to be explored.

\subsubsection{Attention}
The attention mechanism models interactions between words and is a significant component of the Transformer architecture. Enhancements to the attention module have predominantly focused on two aspects: modeling long texts and optimizing the Key-Value (KV) cache.
(1) Modeling Long Texts: The vanilla attention mechanism has a complexity of $O(n^2)$, which significantly increases computational costs for long texts. To address this, Sparse Transformer~\cite{child2019generating} employs sparse attention, utilizing pre-designed attention patterns to avoid the computation of attention over long sequences. Another approach, Reformer~\cite{kitaev2020reformer}, uses Locality-Sensitive Hashing (LSH) to reduce computational complexity. Additionally, \citet{munkhdalai2024leave} compressed context information to shorten sequences, thereby reducing overhead. Others have explored retrieval-based methods~\cite{grave2016improving, izacard2020leveraging}. This area of research continues to hold considerable potential for future advancements.
(2) Optimizing KV Cache: classic Transformers use multi-head attention (MHA), which requires storing extensive key-value caches during inference, slowing down model generation. To mitigate this, \citet{shazeer2019fast} proposed multi-query attention, which employs multiple key heads but only a single value head, substantially reducing the key-value cache and enhancing computational speed. However, \citet{ainslie2023gqa} found that this could degrade model performance, leading to the development of grouped query attention. This method allows multiple key heads to share a single value head, effectively serving as a hybrid between MQA and MHA, balancing computational complexity and performance more effectively. Recently, \citet{deepseekv2} introduced multi-head latent attention, which compresses keys and values into a single latent space, thereby reducing the key-value cache while maintaining robust representational capacity.

\subsubsection{Layer normalization}
Layer normalization (LayerNorm) is important for stabilizing the distribution of hidden states, a key to train large language models. In the classical Transformer architecture, LayerNorm is positioned between residual blocks, hence termed Post-LN. Researchers~\cite{xiong2020layer} observed that this configuration could lead to high gradients near the output layers and very small gradients near the input layers, resulting in unstable gradients and challenging training dynamics.
To address this issue, the Pre-LN configuration was proposed~\cite{xiong2020layer}, placing LayerNorm on the residual pathways before attention or feed-forward network (FFN) module. Experiments have shown that this adjustment leads to more uniform gradient distribution. Building upon Pre-LN, other researchers introduced Sandwich-LN~\cite{ding2021cogview}, which adds an additional LayerNorm at the output of the residual pathways, further enhancing the training stability.
Beyond merely adjusting the position of LayerNorm, researchers have developed DeepNorm~\cite{wang2024deepnet}, which combines a tailored parameter initialization strategy with modified residual connections to stabilize training. This approach enables the training of Transformers with depths reaching up to 1000 layers.
Nevertheless, there still lacks a theoretical understanding about how layer normalization affects the training stability and more work needs to be done for scaling the model even further.


\subsection{Scaling}
\label{sec:foundation_scaling}

Across different information generation scenarios, scaling has been a siginificant factor to the performance improvement. It is largely attributed to the discovery of scaling laws~\cite{kaplan2020scaling}. Scaling laws describe how loss decreases in a log-linear manner as model size or training data volume increases. It can be formulated as follows:
\begin{equation}
L(x)=L_{\infty}+ k \cdot {x}^{-\alpha},
\end{equation}
where $L$ is the loss, $x$ is model size or data size, and $k$ and $\alpha$ are coefficients.
This scaling formula has become a crucial theoretical guide in the era of large models, suggesting that performance can be enhanced at a log-linear rate simply by scaling up the model size or training data.
Based on these scaling laws, researchers also derived optimal model sizes given fixed computational resources~\cite{hoffmann2022training}. Their findings indicate that as computational capacity expands, it is beneficial not only to increase the training step but also the model size. This insight has further facilitated the pursuit of large models.
The correctness of scaling laws was first proposed in language modeling field and then validated in many other areas, including data mixture scaling laws~\cite{ye2024data}, multimodal scaling laws~\cite{henighan2020scaling}, and scaling laws specific to information retrieval~\cite{fang2024scaling}. 

Despite wide recognition of scaling laws, there remains disagreement among researchers about whether scaling is the correct path to the future. This stems from two main concerns: the uncertain relationship between loss and practical metrics, and the inference costs associated with large models.
\begin{itemize}
	\item Loss vs. Metric Improvement: The first arguing point is whether a linear reduction in loss can translate into super-linear improvements in actual metrics. If metrics could improve super-linearly with linear increases in computational effort, scaling up models would be highly advantageous. However, if the decrease in loss only results in linear or sublinear metric improvements, the diminishing improvements make scaling an inefficient option. The relationship between loss and metric performance remains an open question. Some researchers~\cite{wei2022emergent} believe that metrics can improve super-linearly, which is termed emergent abilities. This is further supported by \citet{du2024understanding}, who observed a jump in metrics when loss reaches a certain threshold. Additionally, \citet{power2022grokking} introduced the concept of ``grokking'' to explain emergence, showing that models might suddenly exhibit strong generalization capabilities when provided with sufficient computational resources. Nevertheless, some researchers~\cite{hoffmann2022training} argued that such phenomena do not exist, showing that a well-trained smaller model can outperform a larger, undertrained one. \citet{schaeffer2024emergent} demonstrated that emergent abilities are artifacts of discrete metric functions and found that continuous metric functions do not exhibit such behaviors. \citet{mckenzie2023inverse} even found that scaling results in worse metric scores. The existence of specific emergent abilities remains unresolved and needs to be investigated in future work.
	\item Inference Cost Considerations: Early studies on scaling laws did not account for the higher inference costs associated with larger models. Thus, the arguments that larger models are better~\cite{hoffmann2022training} do not apply when the inference costs are considered. Instead, small models demonstrate potential to lower the inference costs. As shown by \citet{fang2024scaling}, the optimal model sizes become significantly smaller when accounting for inference costs. Besides, \citet{mei2024bigger} show that smaller models can utilize more sampling steps during inference and thus perform better. Consequently, many recent studies focus on extensively training small models. For example, Llama~\cite{touvron2023llama} and MiniCPM~\cite{hu2024minicpm} are trained with data and steps that far exceed the guidance suggested by scaling laws. In the future, the models may be used on a phone to build up intelligent interaction with users. Thus, it is important to develop high-performing small models. 
\end{itemize}

\subsection{Training}
\label{sec:foundation_training}

Generative models in different scenarios are similar in training. For example, they usually use autoregressive training objectives, pretraining-sft-rlhf training stages, and prompt tuning procedure. 
In this section, we focus on the text generation scenario. We first discuss the training objectives and then show the three training stages. Finally, we discuss how to design the prompts after the model is trained. 

\subsubsection{Training Objectives}

For generative language models, the training objective is usually next token prediction. However, this was not widely used when Transformers first appeared. Initially, masked language modeling was the prevalent training objective during the BERT era~\cite{naccl19_BERT}. It masks 15\% of the words in a text randomly, and the model is tasked with predicting these masked words. This approach allows the model to utilize bidirectional attention, enhancing its representational capabilities. Even today, BERT models perform better than autoregressive models on tasks requiring bidirectional attention. However, a significant drawback of this method is the gap between its training setup and downstream tasks, necessitating a fine-tuning phase for adaptation to various applications. Thus, its zero-shot generalization capabilities are very limited.

Next token prediction was developed to address the inability of masked language modeling to generalize zero-shot to downstream tasks. The authors of GPT-2~\cite{openaiblog19_GPT2} proposed that all natural language processing tasks could be reformulated as next token prediction tasks. By training models on this task, models could be directly applied to any downstream task without the need for specific fine-tuning. In fact, research nowadays demonstrates the effectiveness of this idea. Mathematically, next token prediction can be represented with the following formula:
\begin{equation}
P\left(x_{t+1} \mid x_1, \ldots, x_t\right),
\end{equation}
which is to predict the probability of the next token $x_{t+1}$ given the sequence of previous tokens.

\subsubsection{Training Stages}

The training process of language models typically unfolds in three stages: pre-training, supervised fine-tuning (SFT), and reinforcement learning from human feedback (RLHF). Each phase presents unique challenges and methodologies.

Pre-training is the most resource-intensive stage. It is training a randomly initialized model on a large dataset to develop a robust linguistic capability. Several challenges arise during this stage: (1) Large models are especially difficult to train from random initialization. During training, there are often spikes in training loss or difficulty in converging~\cite{zeng2022glm, zhang2022opt, wang2024deepnet}. We discussed various architectural improvements in Section~\ref{sec:foundation_architecture} to address these instabilities, yet a definitive solution remains an open issue. (2) The computational demand is substantial. Pre-training requires stable and efficient use of computational resources~\cite{gpt4}. It often involves parallel processing across multiple machines, which can lead to low utilization rates of computing resources~\cite{chowdhery2023palm}. \citet{zeng2022glm} reported numerous hardware failures during pre-training. (3) The quality of pre-training data is crucial~\cite{gunasekar2023textbooks}. Given the vast amount of data needed, efficiently filtering out low-quality data is essential. The filtering methods usually employ neural scoring models and based on the credibility of the site~\cite{yang2023baichuan, bi2024deepseek}.

Supervised Fine-Tuning (SFT) is to train the model on instruction-response pairs~\cite{chung2022ScalingIL}. The model can thus learns to follow instructions or engage in dialogue~\cite{touvron2023llama}. To enhance dataset diversity, researchers often leverage different types of NLP tasks. The quality of the dataset is significant and requires a skilled annotation team. Besides, it is also important to label safety-related data, which helps instruct the models to learn to reject inappropriate requests~\cite{touvron2023llama}.

Reinforcement Learning from Human Feedback (RLHF) focuses on aligning the model with human preferences based on human feedback~\cite{ouyang2022training, schulman2017proximal}. The process starts by sampling real human prompts to which the model generates multiple responses. These responses are then compared by users or third-party annotators. A reward model is trained based on these human preferences. Subsequently, reinforcement learning techniques utilize the reward model to guide the model updates. This approach significantly enhances the quality of model outputs, especially in creative writing tasks. However, a major challenge is the generalizability of the reward model; as the model evolves, the reward model may no longer accurately assess the quality of outputs. Continuous iterations of this process are necessary to mitigate this issue~\cite{touvron2023llama}. Recently, there are also some offline reinforcement learning algorithms that do not necessaite training a reward model, such as DPO~\cite{NEURIPS2023_DPO}. Yet studies~\cite{xu2024dpo} show that such offline learning methods still underperform the online learning methods.

\subsubsection{Prompt Optimization}

Generative models are highly sensitive to the input prompts; an effective prompt can significantly enhance the quality of the model's output~\cite{liu2023pre}. Therefore, optimizing prompts for a generative model is a crucial area of research. Here are three main directions:
\begin{itemize}
	\item Designing Prompt Templates: Researchers often design prompts that mimic human thought processes to guide the model effectively. This includes using structured thought patterns like chain-of-thought~\cite{wei2022chain}, tree-of-thought~\cite{yao2024tree}, and self-consistency~\cite{wang2022self}, which help the model organize and process information in a logical manner.
	\item Iterative Optimization of Prompt Templates: like reinforcement learning, this method continuously iterate and refine the prompt templates based on the generation feedback. Given that prompt templates are typically discrete, researchers usually employ large language models to conduct prompt updates~\cite{zhou2022large, yang2023large}.
	\item Training Prompt Rewriting Models Using User Interaction Logs: This approach harnesses the rich feedback contained within user interaction logs to tap into user insights. By analyzing how users interact with the model, researchers can train an automated model to rewrite prompts more effectively. This method leverages real-world data to better align the prompts with user intentions and improve the model's responses~\cite{zhan2024capability, zhan2024prompt}.
\end{itemize}

\subsection{Multi-modal Applications}
\label{sec:foundation_mm}
 
The rapid advancement of language models has significantly helped progress in the multimodal domain. Language models facilitate the understanding of multimodal data and developments in multimodal generation. We will discuss these two aspects separately.

\subsubsection{Multi-modal Understanding}

Multimodal Understanding involves models processing inputs from multiple modalities to produce relevant textual responses. For example, GPT-4o can process textual, visual, and auditory input.
The challenges in this area include designing model structures that can handle multimodal inputs and crafting appropriate training objectives. Here, we focus on how visual signals are integrated into large language models:

In terms of aligning multimodal inputs, there are mainly three approaches:
\begin{itemize}
	\item Object Detection-Based Input: This method involves detecting objects within an image, extracting their features and associated spatial information, and then feeding this data into the language model~\cite{lu2019vilbert, chen2020uniter}. While this approach is effective, it tends to be slow due to the processing time required for object detection.
	\item Visual Encoding: Another method encodes images directly using a visual encoder, which converts images into a latent vector representation before integration with the model~\cite{huang2020pixel, wang2022ofa, alayrac2022flamingo, wang2023cogvlm, li2023blip}. This method can sometimes result in the loss of detail.
	\item Patch-Based Input: The most efficient approach involves dividing images into several patches, transforming them with a simple linear layer, and directly inputting them into the model without the need for a complex visual encoder~\cite{kim2021vilt}.
\end{itemize}

In terms of training methods, there are mainly four types of training objectives:
\begin{itemize}
	\item Contrastive Learning or Image-Text matching: These tasks require the model to correctly categorize images and their corresponding textual descriptions, aligning the representations of text and images~\cite{li2021align, radford2021learning, li2023blip}.
	\item Image Captioning: The model generates captions based on images, which helps it learn to understand the visual content~\cite{alayrac2022flamingo, wang2022ofa, wang2023cogvlm, li2023blip}.
	\item Fine-Grained Image Understanding: The model is tasked to describe specific areas of an image or locate particular objects within an image. This helps enhance the model's detailed comprehension of visual elements~\cite{wang2022ofa, yu2023rlhf}.
	\item Image Generation: This task is reconstructing the original pixels of an image that has been blurred or corrupted~\cite{wang2022ofa, bao2021beit}.
\end{itemize}

These methodologies and training objectives are crucial for advancing models' capabilities to process and interpret complex multimodal information effectively. This facilitate a more natural interaction with users.

\subsubsection{Multi-modal Generation}

Multi-modal generation models, such as text-to-image generation, have substantially revolutionized the field of art creation. Traditionally, GAN~\cite{reed2016generative} and autoregressive methods~\cite{ramesh2021zero} are mainstream methods. However, they are computationally expensive and can not produce high-quality results. Recently, diffusion~\cite{nichol2021glide, rombach2022high} emerges as a new state-of-the-art method in multimodal generation. It perturbs the data with noise and learns to reconstruct the original data. 

Language models are increasingly applied in the multimodal generation domain, such as in image~\cite{ho2020denoising, zhang2023text} and video generation~\cite{peebles2023scalable, singh2023survey}. Language models are primarily utilized for processing training data and reformulating prompts.

In terms of training data, the titles associated with real-world images or videos often contain significant noise. If generative models are trained directly on these noisy titles, it could lead to inaccurate semantic understanding. To address this, language models can be used to filter and regenerate text descriptions within the training data~\cite{betker2023improving, videoworldsimulators2024}. For instance, a multimodal understanding model could first be trained, then used to relabel videos or images to obtain more precise and detailed text descriptions. Experimental results have shown that this method significantly improves the fidelity of model generations to prompts.

During inference, multimodal generation models are highly sensitive to the input prompts. Many users do not know how to craft effective prompts and thus get unsatisfying responses~\cite{oppenlaender2023taxonomy}. As a result, it is common to train a language model to rewrite user-provided prompts to enhance the quality of the generated images~\cite{betker2023improving}. One of the challenges here is the difficulty in annotating such rewriting training data, as even system developers may not always know the optimal prompts, let alone crowd-sourced workers~\cite{liu2022design}. To overcome this, some researchers collect a large number of user-shared effective prompts as training data~\cite{hao2024optimizing}. Others build prompt-rewriting models based on user log data, capturing preferences and feedback for training~\cite{zhan2024capability}.

\section{Information Synthesis}\label{ch2-syn}




Other than generating information directly, another important research and application direction is to use the power of generative AI models, particularly LLMs, to integrate existing information and generate grounded responses accordingly.
For simplicity, we refer to this paradigm as \textit{information synthesis}.
The key difference between information generation and information synthesis is the source of information.
Information generation relies on the internal knowledge and information gathered through the training of generative AI models to create the model outputs, while information synthesis requires external sources to provide information to the models, and the models serve more as a integrator than a creator.
There are multiple reasons why information synthesis is considered more reliable than generation in several IA scenarios. 
Here we discuss two of the most significant ones, i.e., model hallucination and external knowledge.

Hallucinating, which refers to the behavior of generative AI models that create responses and outputs that are not grounded by facts or existing supporting materials, is rooted in the foundation of most existing generative AI systems.
For instance, LLMs create responses based on the next token prediction task, which formulates the generation of language as a probabilistic process and generates the next token in the output based on a probabilistic distribution (over the vocabulary) predicted by neural networks~\cite{gpt4,touvron2023llama}.
The probabilistic model of LLMs allows them to capture knowledge in large scale data efficiently and effectively, but it also introduces inevitable variance in their generation process.
In other words, it is well acknowledged that it's theoretically impossible to prevent LLMs from generate data that are not seen in their training process~\cite{ji2023survey}.
While the ability of hallucinating is the source of creativity for LLMs (and for human as well), it's not always desirable in practice, particularly for tasks with high requirements on result precision, reliability, and explanability.
Therefore, asking the generative AI models to integrate human created or factually grounded materials instead of generating information on their own is often considered more effective and robust to hallucination-sensitive applications.

The need of external knowledge is another key reason why we may prefer information synthesis over information generation.
Despite the fact that modern generative AI models are trained with incredibly large amount of data gathered from the Web, there are many cases where we still need to retrieve and find supports from external knowledge collections to finish certain tasks.
Examples including the use of private datasets, vertical domain applications that require special knowledge, tasks that involve time-sensitive data, etc.
It is usually inefficient or prohibitive to update large-scale generative AI models such as LLMs with task-oriented external data through model pre-training or supervised fine-tuning (SFT)~\cite{arefeen2024leancontext, aharoni2020unsupervised,li2024blade}.
Even if possible, such paradigm is not preferred because the internal knowledge structures of most generative AI models are still mystery (at least of today), and there is no guarantee that the models could behavior and use the external information as we expect.
In contrast, using generative AI models as information synthesizer gives us not only more flexibility, but also more transparency and control over system outputs.

In this section, we discuss how generative AI models, particularly LLMs, can serve as effective information synthesizers for IA. 
We start with introducing one of the most popular information synthesis paradigm, i.e., retrieval augmented generation (RAG), and then discuss several other directions that utilize LLMs for corpus modeling and understanding.

\subsection{Retrieval Augmented Generation}\label{ch2:RAG}

Retrieval Augmented Generation, or RAG, refers to the process of augmenting LLMs with data retrieved from external collections or synthesizing multiple retrieval results with LLMs for downstream applications~\cite{gao2023retrieval,lewis2020retrieval}.
While the popularity of RAG rose after the release of large-scale pre-trained language models such as GPT~\cite{gpt4} and BART~\cite{lewis2020bart}, relevant topics and techniques have already been studied for at least more than two decades in both IR and NLP communities, e.g., extractive and abstractive summarization that generates summary based on retrieved sentences~\cite{moratanch2017survey,lin2019abstractive} or answer extraction from top retrieved document~\cite{bajaj2018ms}.
A major reason why RAG-like techniques were not as attractive as they are today is the limited performance of generative models before the era of LLMs.
After ChatGPT~\cite{gpt4} demonstrated superior ability text generation at the end of 2022, there have been many studies and surveys on RAG and its applications in LLMs~\cite{gao2023retrieval,zhao2024retrieval,asai2023retrieval}.
As the intent of this chapter is not to provide yet another survey on existing RAG papers, we focus the following discussions on several present and future directions for RAG and their relations underneath.

\subsubsection{Naive RAG}

Naive RAG refers to the paradigm that directly feeds documents or other types of information retrieved by a retrieval system to the input (e.g., prompts) of a generative AI model and hope that the model can generate better output with or without a specific target task~\cite{guu2020retrieval}.
It is also referred to as the ``Retrieve-then-Read'' framework that has been used in reading comprehension and text summarization before LLMs hit the world~\cite{ma2023query}.
Given an input (could be a query or a specific task instruction), we first retrieve relevant information (usually entities, passages, or documents) from an external corpus or previous inputs (e.g., the memory of an agent~\cite{wang2024survey,zhang2024survey}) with a retrieval system.
Then, we craft a input prompt with the retrieval results and feed it to the LLM.
The LLM will generate the final response based on the input request and the retrieved information.
This paradigm has already been proven to be effective in multiple IA tasks such as question answering~\cite{lewis2020retrieval}.

Because LLMs are purely used as black-box tools to process the retrieved documents and input request in naive RAG, existing studies on this direction mainly focus on the development of better retrieval systems and prompt design for RAG.
The studies on retrieval systems, unsurprisingly, are highly similar to those in IR, which involve indexing, query processing, first-stage retrieval, re-ranking, etc. 
These topics and system components have already been studied in the IR community for more than five decades.
Perhaps the most notable difference is that recent studies on naive RAG often prefer the use of neural retrieval models (e.g., dense retrieval models~\cite{zhan2021optimizing}) over traditional term-matching models (e.g., BM25~\cite{FIR09_BM25}). 
An important reason behind this is that neural retrieval models share similar theoretical background and model structures with LLMs.
This makes joint optimization possible in modern RAG systems, which we discuss in Section~\ref{sec:joint_RAG}.

The design of input prompts with retrieval results, on the other hand, is relatively more under-explored before the rise of LLMs.
It has been well recognized that prompt formats, even when the contents are same, could significantly affect the performance of LLMs.
How to feed retrieval results effectively into the prompts of LLMs for RAG has thus attracted a lot of attention recently~\cite{ma2023query,mao2024rafe,chan2024rq}.
Studies have found that LLMs exhibit significant position bias over the input result sequences~\cite{li2024long,liu2024lost}, and has different perspectives on relevance with human experts~\cite{faggioli2023perspectives}.
Since prompts are the main interaction interface between retrieval and generation, their design principles and downstream effects on naive RAG are of great value both in research and real-world applications.
Particularly, how to craft effective RAG prompts automatically could be a fruitful direction to explore.
Existing studies have shown that high-quality prompt writers can be automatically learned based on downstream task performance and user logs in image generation~\cite{zhan2024capability}, and it is widely believed that similar techniques have also been used in popular LLM chatbots~\cite{10.1145/3560815}.
Yet, how to do this for RAG remains to be a question to be answered.

\subsubsection{Modular RAG}

In contrast to naive RAG methods, modular RAG treats retrieval systems as functional modules to support LLMs~\cite{wang2023knowledgpt}.
While some works view this retrieval module as one type of many tools that can be learned and used by LLMs~\cite{qin2023tool}, it is widely acknowledged that retrieval systems possesses a irreplaceable position in modern LLM applications due to its diverse nature and significant importance~\cite{gao2023retrieval}.
Broadly speaking, existing studies on using retrieval systems as functional modules for LLM generation mainly focus on the three ``W'' questions, namely \textit{when to retrieve},\textit{what to retrieve}, and \textit{where to retrieve}.

The question of \textit{when to retrieve} refers to the timing of functional call for retrieval systems. 
In contrast to LLMs that directly create responses based on their internal parameter space without explicit evidence grounding, retrieval systems produces reliable and explainable information directly by searching external corpus.
From this perspective, the best timing to call the retrieval system is when LLMs start to hallucinate or produce wrong results. 
Yet, identifying such timing is difficult because we neither know the correct answers in advance or understand the internal mechanism of LLMs (at least of today)~\cite{jiang2023active}.
One naive yet effective method is to retrieve supporting evidence for LLM inference with a fixed time interval, such as every fixed number of generated tokens\cite{ram2023context, borgeaud2022improving} or every sentence~\cite{trivedi2022interleaving}.
More advanced paradigms involve the analyze of knowledge boundary~\cite{ni2024llms} and the estimation of prediction uncertainty in LLMs~\cite{jiang2023active,su2024dragin}.
Theoretically speaking, since the study of \textit{when to retrieve} shares similar motivations and foundations with the study of hallucination detection, existing studies on LLM hallucination~\cite{su2024unsupervised, liu2021token} could provide important inspiration for research on this topic. 
Promising directions including better fact checking systems for LLMs~\cite{fadeeva2024fact} and more investigations on how to characterize the confidence and uncertainty of LLM predictions based on both external behavior and internal state analysis~\cite{su2024dragin}.

\begin{figure}[ht] 
	\centering
	\includegraphics[width=0.7\textwidth]{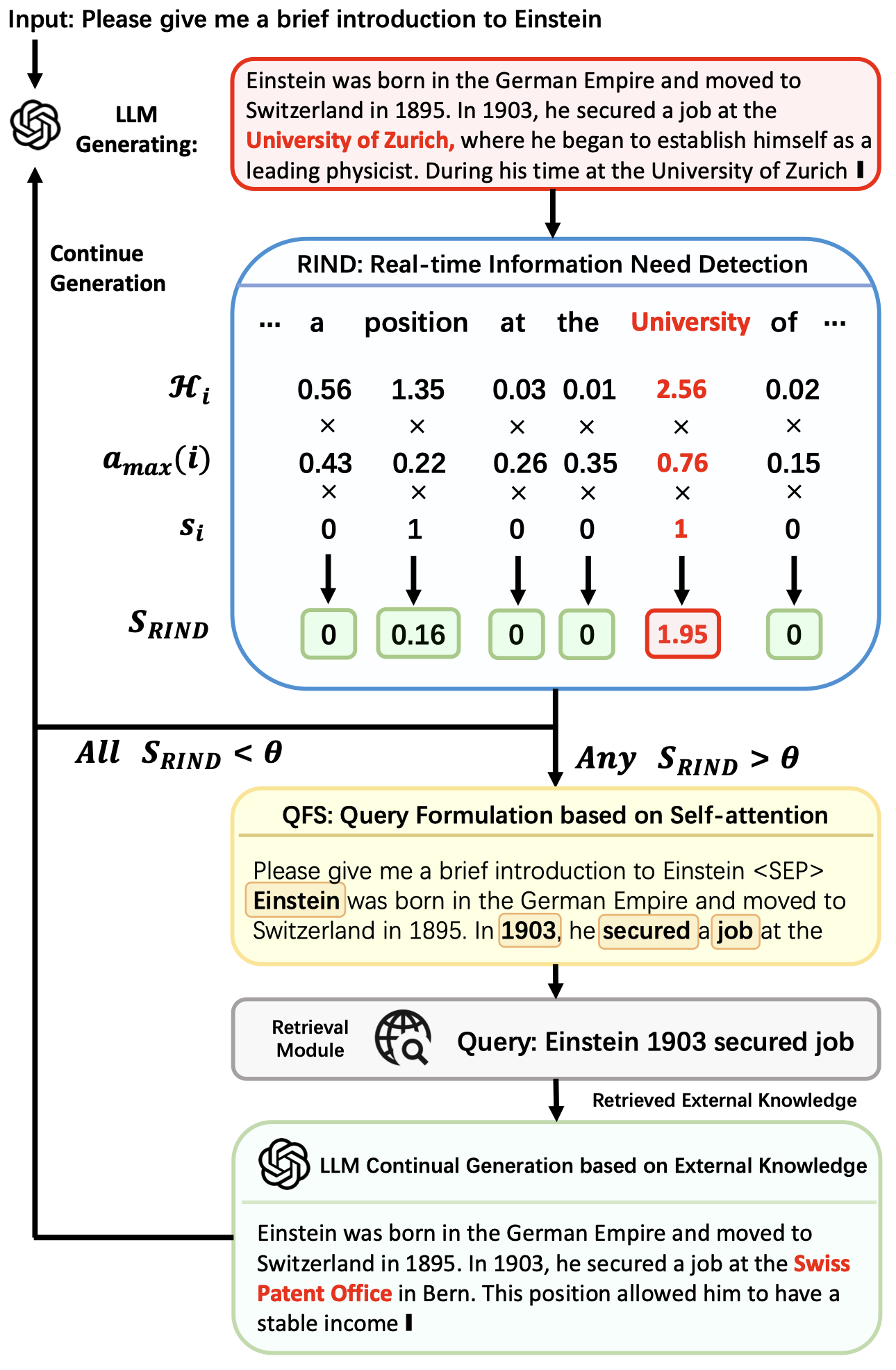} 
	\caption{Su et al.~\cite{su2024dragin} generate queries for RAG based on the internal attention distribution of LLMs.}
	\label{fig:dragin}
\end{figure}

The question of \textit{what to retrieve} focuses on analysis the intents and information needs of LLMs in inference.
LLMs often need the help of different tools and systems to finish different tasks~\cite{qin2023tool}.
However, in contrast to other tools widely studied in tool learning, retrieval itself is a complicated systems with dynamic and free-form inputs, data collections, and outputs. 
Therefore, understanding what exactly is needed by LLMs and how to formulate it in the language of retrieval systems is an important problem.
Most existing studies on RAG naively use the whole or local context of LLM inference as the queries to retrieval systems and assume that these context contain enough information to guide retrieval~\cite{zhao2024retrieval}.
A slightly better solution is to use the terms that LLMs have low confidence to formulate queries since uncertain tokens represent cases where LLMs have limited knowledge to generate responses and thus need more information~\cite{jiang2023active}.
As long studied in the IR community, the formulation of an effective query requires deep understanding of the user's intent, and many of the important context information behind a user intent is not explicitly expressed in the words they wrote~\cite{cronen2002quantifying}.
Therefore, a more theoretically principled method to answer \textit{what to retrieve} in RAG is to analyze the internal state of LLMs and infer their information needs directly.
For example, Su et al.~\cite{su2024dragin} directly formulate queries based on the internal attention distribution of LLMs (Figure~\ref{fig:dragin}) and improve the performance of RAG for nearly 20\% on several benchmark datasets without changing the retrieval system. 
This demonstrates the potential of future studies on this direction.

\textit{Where to retrieve} refers to the question of how to identify the correct information sources for RAG. 
Studies on this direction is particularly related to the research on multi-source retrieval~\cite{doi:10.1142/S0218215793000071} and tool learning~\cite{qin2023tool}.
To answer different requests related to the use of information collected from different databases or data collections, LLMs need to learn how to interact with each information sources effectively and efficiently. 
The studies of tool learning focus on teaching LLMs to use tools according to the context, and retrieval systems are usually considered as one type of tools to use.
However, retrieval itself could be a complicated problem when we possess multiple data collections with different characteristics.
In search engines, information sources are broadly categorized based on their modality, and we usually build separate systems for each of them (e.g., the ``Images'', ``News'', ``Videos'' tabs on Google). 
While commercial search engines may aggregate results from different sources into a single page, the ultimate search engine result page (SERP) shown to users are just a list of results and it's up to the users to decide which they want to see and how to use these results for downstream applications.
In contrast, when using LLMs, users often request LLMs to directly answer their question instead of listing a couple of candidates~\cite{wang2024usercentric,wang2024understanding}, so it's the job of LLMs to decide where to retrieve the information given the current context.
While the studies of how to navigate user queries to search indexes built from different information sources have been widely studied in the IR community~\cite{beitzel2004fusion,wu2006performance,cormack2009reciprocal,lee2015optimization}, how to do it for RAG with modern generative AI models is, to the best of our knowledge, still underexplored. 
Existing literature on RAG mostly works on a single retrieval collection (usually a text corpus), but it's obvious that no single collection can satisfy the needs of LLMs in different tasks.
For instance, when writing a legal case document, the judge needs to collect and organize information from evidences, complaints, counterclaims, court records, as well as legal articles and previous cases. 
How to navigate the generation model to retrieve and integrate information from different sources jointly for downstream applications is a practical and potentially fruitful research question for RAG.

\subsubsection{Optimization of Retrieval and Generation}\label{sec:joint_RAG}

As discussed in several RAG surveys~\cite{gao2023retrieval,zhao2024retrieval}, the optimization of RAG systems usually involves the optimization of three components, i.e., the retriever, the generator, and the augmentation method.
If we further step back and look at the high-level goals of RAG optimization, we could also categorize it based on how we evaluate the RAG system, namely the evaluation from the perspectives of retrievers, generators, or the joint systems.
The evaluation from the retriever perspectives is not particularly different from existing studies on ranking evaluation. 
The underlining assumption of this is that, once the LLMs are fed with the passages or documents that contain the correct information, they should be able to produce the correct answers directly.
Therefore, the evaluation and optimization of a RAG system could downgrade to the evaluation and optimization of a classic retrieval/ranking systems, to where most existing works on dense retrieval and LTR could be applied~\cite{liu2009learning,zhan2020learning}.
Yet, there are still differences between RAG and traditional retrieval tasks as the queries are no long issued by users.
How to formulate queries efficiently and effectively from LLMs for the retriever is worthy research question, and studies on this direction has already shown potentials in improving the overall quality of RAG systems~\cite{su2024dragin}.

From the perspective of generators, RAG evaluation and optimization focus more on improving the robustness and effectiveness of LLM generation based on a fixed set of retrieval results~\cite{borgeaud2022improving}.
This often means extra training or fine-tuning on LLMs to improve their fundamental ability in information processing.
For example, retrieved documents could be lengthy, and LLMs are usually not good at processing long input context~\cite{liu2024lost}.
Therefore, how to design efficient LLMs that can take long context inputs efficiently and effectively has been a popular research problem that have been widely studied by researchers from both academia and industry~\cite{li2024long}.
We have seen many companies show off their models based on how many input tokens they can process in one request.
In addition, since retrieval results are fed as a part of the LLM inputs, whether the LLMs can generate the response based on the retrieved documents instead of their internal knowledge could be seen as a special type of instruction-following ability.
Studies have been conducted to teach LLMs to utilize retrieval results faithfully and constantly in RAG systems~\cite{arora2023gar}
On the other hand, factors such as irrelevant results and ranking perturbations are well acknowledged to be harmful for the performance of generators in RAG, so there are also studies that try to improve the robustness of LLMs from the perspective of RAG.
For example, \citet{zhang2024raft} proposes to fine tune LLMs with the presence of retrieval results (i.e., retrieval augmented fine tuning) so that LLMs can learn the domain-specific knowledge introduced by the retriever and improve their robustness against potential distracting information from retrieval.

From the perspective of augmentation methods, existing research mostly focuses on the joint optimization of RAG system as a whole.
In other words, the loss functions of RAG optimization should be built from the performance metrics of downstream tasks directly.
While this paradigm is appealing, it often has strict requirements on the design of RAG systems.
Particularly, it's difficult to apply such joint optimization algorithms on a RAG system in which retrievers and generators are loosely connected through prompts constructed from discrete retrieval results.
While reinforcement learning could solve the problem in theory, its empirical performance when being used as the solo optimization algorithms for ranking systems is still not satisfying at this point~\cite{xu2022reinforcement}. 
If you already have a good retriever and only conduct fine-tuning with a fixed LLM, then it may work~\cite{shi2023replug}, but this still doesn't look like a perfect solution because reinforcement learning usually subject to large variance in practice.
To the best of our knowledge, how to directly connect the training of retrievers with the auto-regressive loss of the generators in RAG is still an open question.
Answering this question requires us to go deep into the structure of generative AI models and retrieval models, and develop new model structures that can take advantages from studies on both sides.



%

\subsubsection{Retrieval Planning and Composite Information Needs}

As discussed above, the initial motivation behind the studies of RAG mostly focuses on using the power of retrieval systems to improve the quality of responses generated by LLMs in terms of reliability and informativeness.
While it is widely acknowledged that problems such as hallucination and high computation cost in supervised fine-tuning will continue to be significant for generative AI models in a short period of time, there are also concerns, especially from the IR community, that retrieval could become less important with the rapid evolution of LLMs~\cite{ai2023information}.
In fact, ChatGPT has already shown similar accuracy and better user satisfaction on factoid question answering than traditional web search engines~\cite{gpt4}.
However, the rise of generative AI models also brings brand new opportunities for IR. 
One of them is the possibility of moving from SERPs that simply list result candidates to a real information agent that solve complicated tasks with composite information needs.

Today, most people treat IR systems as \textit{unit information solvers}.
Despite of their actual task characteristics, users first decompose their goals into a couple of unit information need (usually expressed with separate queries), and then issue them one by one to search engines or recommendation systems to find the corresponding answers.
An important reason behind the popularity of this paradigm is that, at least of today, IR systems are not capable of doing complicated information tasks with composite needs and multi-step planning.
For example, we can use a search engine to find a survey on RAG by searching "survey of RAG", but cannot write such a survey directly by retrieving and analyzing papers from publication collections.
The job of information need decomposition and retrieval planning has always been human's.

Fortunately, with the help of generative AI models like LLMs, it is now possible to push the boundary of IR systems and tackle such advanced information tasks for users.
Composite retrieval is not a new concept in IR~\cite{bota2014composite}, but previous studies refer to the phrase as retrieval paradigms that cluster results from multiple sources and show them in groups for specific user queries~\cite{amer2014composite}. 
While this represents one type of composite needs, it is relatively simple as the target user queries usually are mostly topic-specific and keyword-based.
Complicated information tasks such as survey generation and professional document writing often involve multi-step planning and multi-round interactions between the retrieval results and response generation.
To build powerful IR systems or agents that can solve such composite information tasks, we need to construct collaborative systems that deeply connect the retrieval, planning, and generation.
For instance, we need to conduct generation-oriented retrieval optimization to build retrieval framework and model interfaces for downstream task planner and response generators; we also need to design retrieval-oriented generation models that can decompose information needs, navigate the retrieval process, gather information from multiple sources to generate the final results.
Research on these directions could be fruitful and significantly extend the scope of IR in the era of generative AI. 

\subsection{Corpus Modeling and Understanding}\label{ch2:corpus}

In contrast to using RAG, another line of studies try to use generative AI models to replace traditional retrieval systems.
Directly answering user's information need instead of showing ten blue links has long been an important goal for the development of intelligent IR systems~\cite{kolomiyets2011survey}.
With the rise of LLMs, such vision is now achievable in a significant extent.
For example, LLM-based chatbots like ChatGPT can answer multiple types of user queries with direct answers~\cite{wang2024understanding}.
\citet{10.1145/3476415.3476428} has discussed several paradigms in which pre-trained language models can help IR systems answer user's information needs directly without listing references.
The intuition is to use neural network based language models to store the corpus knowledge in parameter space and pull relevant answers or information directly from it based on user's queries.
Depending on how the problem is formulated, several research directions have emerged.
Specifically, in this section, we discuss two of them, namely generative retrieval and domain-specific modeling.

\subsubsection{Generative Retrieval}
The idea of \textit{Generative Retrieval} comes from the idea of differentiable index proposed by \citet{10.1145/3476415.3476428}.
The original name used in the paper was \textit{Model-based IR}, but after the rise of generative AI models, some researchers start to refer to studies on this direction as generative retrieval (GR).
The core idea of GR is two-fold, i.e., the differentiable index and the generation of doc IDs.

Inspired by the superior performance of pre-trained language models, particularly BERT~\cite{naccl19_BERT} and GPT~\cite{gpt4}, generative retrieval wants to explore the possibility of replacing traditional term-based index (e.g., inverted index) in retrieval systems with large-scale neural networks.
In contrast to dense retrieval models that build neural encoders to project documents to latent semantic spaces and build explicit indexes based on document vectors, GR tries to build implicit indexes in the parameter space of neural networks.
For instance, DSI and its variations~\cite{zhuang2022bridging,tay2022transformer,tang2023semantic,sun2024learning} have tried to train pretrained language models on the target corpus directly and then treat the model's parameter as a ``index'' of the corpus.
Studies on this direction argue that, by training the neural models to encode the whole corpus, documents and information would be implicitly stored in the parameters of the models, and this parameter-based indexes have better storage efficiency than traditional term-based or vector-based indexes~\cite{tay2022transformer}.
They also argue that such paradigm can unify the multi-stage retrieval pipeline so that indexes can be trained directly for the final retrieval objectives.
However, storing raw document content directly in limited parameter spaces often lead to significant information loss (which is reflected in the suboptimal retrieval performance of GR models~\cite{nguyen2023generative}), and using model parameters as indexes make the whole system uncontrollable by both system developers and users.
While the former could be alleviated by using large-scale models, the later is still an unresolved problem for GR.
For example, it's difficult, if not impossible, to remove or update a document indexed in the parameter space when we don't know what exactly each parameter do in the neural models.
Considering that dense retrieval models built with product quantization and inverted file systems can achieve state-of-the-art retrieval performance with similar latency and less storage than term-based models with inverted indexes~\cite{zhan2022discrete}, whether the idea of differentiable indexes in GR worth its price is still a controversial question.

Another important characteristic of GR models is to retrieve documents by generating sequences of doc IDs through auto-regression.
Because documents are stored implicitly in model parameters, to actually retrieve a real document, GR models use user's queries as prompts to generate document IDs, which usually consist of a couple of special tokens, that exclusively identify each relevant document.
Since the birth of GR, a variety types of document IDs have been proposed, which can be broadly categorized as IDs with explicit tokens~\cite{zhuang2022bridging,tay2022transformer,tang2023semantic} and IDs with implicit tokens~\cite{10.1145/3589334.3645477,zeng2024planning,sun2024learning}.
GR models with explicit ID tokens try to label each document with sequences of real terms that have semantic or numerical meanings. 
Examples including keyword-based doc IDs and tree-based doc IDs~\cite{tay2022transformer}.
Compared to vectors in dense retrieval, these methods have less flexibility and capability in document modeling as they discretize document semantic meanings with limited number of tokens, and their retrieval performance is usually poor~\cite{10.1145/3589334.3645477}.
However, they have better explanability than other neural retrieval models because their doc ID tokens are constructed from real words or document clusters.
To avoid the theoretical limitation of explicit token IDs and grant GR models with the same modeling capacity of dense retrieval models, several studies have proposed to build implicit token IDs with latent vectors~\cite{10.1145/3589334.3645477,zeng2024planning,sun2024learning}.
The idea is to represent each document with a sequence of latent vectors so that fine-grained semantic information would not be lost.
These types of GR models are highly similar to existing dense retrieval models since both of them represent each document with latent vectors.
The major difference is that the former uses a sequence of vectors from a learned codebook constructed in training, while the later builds separate vectors for each document directly from their raw content.
\citet{wu2024generative} have proved that the GR models with implicit tokens are equal to a multi-vector dense retrieval models in theory.
Also, the use of a learned codebook for implicit token vectors is theoretically the same with a dense retrieval system that uses cluster-based product quantization~\cite{zhan2021joint,zhan2022discrete}.
Therefore, the performance upper bound of GR (with implicit tokens) and dense retrieval is the same in theory.
While some believe that GR models could have lower latency as they don't need to search among millions of documents on the fly, this is a questionable argument because the inference of a large-scale neural model is usually much slower than a vector-based search on distributed systems.
Also, the maintenance of information in a neural model is much more difficult than a vector-based database. 
Perhaps the future potential of GR does not lay in retrieval effectiveness or efficiency, but some other perspectives such as explainability. 


\subsubsection{Domain-specific Modeling}

LLMs, particularly those with instruction tuning, can response to user's queries directly.
This exactly matches the initiative of a long-standing vision of IR systems to directly answer user's need without listing a couple of documents~\cite{10.1145/3476415.3476428}.
Therefore, ever since the rise of ChatGPT, there has been serious discussion on whether LLMs are future seach engines in practice~\cite{ai2023information}.
Yet, apart from the hallucination problem discussed in previous sections, there are other challenges that prevent generative AI models like LLMs to serve as a major information accessing tool for modern users.
One of them is how to teach LLMs to understand and use knowledge from external corpus not included in their initial training process.
If we treat each external corpus as a domain-specific dataset, then the studies on this direction is essentially the same with the construction of domain-specific LLMs.
While RAG can help LLMs adapt to new domains quickly, their performance is limited when the understanding of input documents from the external corpus requires domain knowledge that the LLMs do not possess in advance~\cite{li2024blade}.

To solve the above problem and build usable IA systems with LLMs on domain-specific data, one of the most popular method is to conduct continue pre-training or supervised fine-tuning of LLMs on the target domain corpus. 
The idea is to apply similar training strategies used in model pre-training on the new corpus so that LLMs can better capture knowledge in the new domain.
Example studies on this direction include techniques on data selection~\cite{aharoni2020unsupervised} and tokenizers adaptation~\cite{sachidananda2021efficient} that directly use the target corpus to train LLMs.
Many domain-specific LLMs have been developed, include legal LLMs, financial LLMs, etc.~\cite{huang2023lawyer,cui2023chatlaw,wu2023bloomberggpt}
The continue pre-training of LLMs on external corpus has been shown to be effective on many domain-specific tasks such as domain QA and text generation.
However, modeling external corpus through this method may not be preferred in practice when we don't have enough computation resources to train LLMs or can't access the parameters of them.
Also, till the end of the today, the internal knowledge structure and learning mechanism of LLMs are still unknown, and applying naive continue pre-training algorithms on external corpus could hurt the performance of LLMs in unexpected way.
Therefore, researchers have designed several knowledge editing techniques on LLMs to explore the possibility of injecting knowledge with no or low cost on the general effectiveness of LLMs~\cite{dai2021knowledge,meng2022locating}.
Studies on this direction is still in an early stage as most existing methods only work on fixed and limited updating rules and knowledge entity triples~\cite{liu2024evedit}, but it could be fruitful in future since domain adaption and external corpus modeling is a wide need of LLM applications in practice.

Besides continue pre-training, another paradigm to model external corpus and domain knowledge is to build separate language models for each corpus and combine them with the large general LLMs to form a collaborative system.
The intuition behind this is relevant to the idea of LLM agents where each LLM could serve as different roles in the system to accomplish tasks together.
It is widely acknowledged that the phenomenon of emergent abilities only present in large-scale models~\cite{wei2022emergent}, but the training cost of such models (e.g., GPT-4~\cite{gpt4}) is usually prohibitive to research institutes and small companies, even with parameter efficient algorithms~\cite{hu2021lora}
Inspired by the superior instruction following ability of LLMs, researchers have explored the possibility of building small models for external corpus modeling and use them to communicate domain-specific knowledge to large general LMs~\cite{li2024blade}.
In other words, the small models can serve as domain knowledge `` consultants'' and the large general models can serve as the decision makers that finish domain-specific tasks based on the guidance of the small models.
Experiments have shown that such paradigm can improve black-box LLMs' performance on domain-specific tasks with low cost and high flexibility.
While the overall idea of prompt general LLMs with domain-specific prompts is similar to the framework of RAG, building an actual LM for corpus modeling enable us to capture implicit domain knowledge (e.g., the fine-grained differences between law articles~\cite{li2023sailer}) and potentially save tokens in prompts. 
There are concerns on whether this paradigm is still worthy when we have more powerful LLMs that include more domain-specific data in training.
However, since many users prefer to keep their data private to themselves due to multiple safety and privacy concerns, this paradigm and RAG could continue to be appealing in practice.



\section{Summary and Future Directions}

In this chapter, we introduce the foundations and applications of generative AI models in information accessing.
Instead of analyzing how generative AI models like LLMs could improve the existing modules of search engines and recommendation systems, we focus on how the they could revolutionize information access with new methodologies and system design. 
Particularly, we discuss two new paradigms brought by generaive AI models, namely information generation and information synthesis.

Information generation refers to the scenarios where users can use generative AI models to create information that directly satisfies their information needs.
Here, we delved into the core components of generative models, including model architectures (with a focus on Transformers and their improvements), scaling laws, and training methodologies. 
We examined the debates surrounding continual model scaling, the importance of prompt optimization, and the extension of these models to multi-modal applications for information access.

Information synthesis refers to the paradigm that utilizes the superior instruction-following and logic-reasoning ability of LLMs to aggregate and synthesize existing information.
We extensively discuss one of the most representative techniques, i.e., Retrieval Augmented Generation (RAG), on this direction, and introduce various approaches from naive implementations to more sophisticated modular systems. 
We describe the challenges and opportunities in optimizing RAG systems, highlighting the need for joint retrieval-generation optimization and the potential of several relevant research directions such as composite retrieval with planning.
Besides RAG, we also discuss some alternative paradigms that use generative AI models to model corpus knowledge directly, such as generative retrieval, which aims to replace traditional indexing methods with neural network-based approaches, and domain-specific model training, which conducts continue pre-training or fine-tuning on LLMs with the target corpus.
We discussed the potential and limitations of these approaches, including issues of system controllability and cost efficiency.


Overall, research on how generative AI models could reshape modern information access systems is still at an early stage today.
As discussed above, existing studies on information generation and information synthesis either focus on simple information tasks (such as writing a poem, answering a factoid question, etc.) or reply on simple system design (e.g., feeding all documents to LLMs as prompts) that obviously cannot fully exploit the power of modern retrieval and generation models.
Therefore, we believe that there are two major directions worth exploring in the next couple of years (at least).
The first one is to move from simple and unit information retrieval tasks (e.g., factoid QA) to more complicated information tasks that used to be ``impossible'' for modern IR systems.
Examples include retrieval with composite needs (e.g., "help me plan a wedding in Amherst, MA") or tasks that requires planning and multiple rounds of retrieval and generations (e.g., "write a survey on RAG").
These tasks used to require human experts to decompose the needs and conduct retrieval, analysis, and result aggregations.
With the help of generative AI, accomplishing them automatically with machines is now possible. 
The second direction is to explore better techniques to communicate, collaborate, or even unify retrieval and generation systems for information accessing.
While the studies of RAG have attracted considerable attention, existing works mostly use retrieval systems as plug-in tools for LLMs without digging into their internal connections and differences.
Examples such as how to understand the information needs of LLMs, how to communicate the retrieved results to LLMs, and how to optimize generators for retrieval and retriever for generation are all important yet underexplored research topics.
There are many questions related to each of these topics that worth detailed investigation, including the design of new training paradigms, the development of agent-like system frameworks, potential problems and bias introduced by off-policy and on-policy training for the joint system, etc. 

When ChatGPT first arrives, there are people from the IR community worried that such generative AI models could overthrow all existing IR systems and crush everything in the field~\cite{ai2023information}, as it has almost happened in NLP.
Interestingly, in simulated social experiments on human-AI competitions, \citet{yao2024human} find that, if human producers don't extend their capacities with the help of generative AI, they will eventually be ``replaced'' by AI.
From this perspective, the future of IR research in the era of generative AI lies in how to extend the scope of IR with generative AI models to finish more complicated information tasks and develop more general system architectures that not just retrieve a list of document, but perform more sophisticated information processing and planning.





\bibliography{sn-bibliography}

\end{document}